\documentclass[10pt,conference]{IEEEtran}
\IEEEoverridecommandlockouts
\usepackage{cite}
\usepackage{amsmath,amssymb,amsfonts}
\usepackage{algorithmic}
\usepackage{graphicx}
\usepackage{textcomp}
\usepackage{xcolor}
\usepackage{balance}
\usepackage[hidelinks]{hyperref}
\def\BibTeX{{\rm B\kern-.05em{\sc i\kern-.025em b}\kern-.08em
    T\kern-.1667em\lower.7ex\hbox{E}\kern-.125emX}}
\begin{document}

\title{ICVul: A Well-labeled C/C++ Vulnerability Dataset with Comprehensive Metadata and VCCs\\
}

\author{\IEEEauthorblockN{Chaomeng Lu, Tianyu Li, Toon Dehaene, Bert Lagaisse}
\IEEEauthorblockA{\textit{DistriNet Group-T, KU Leuven}\\
Leuven, Belgium \\
\{chaomeng.lu, tianyu.li, toon.dehaene, bert.lagaisse\}@kuleuven.be}}

\maketitle

\begin{abstract}
Machine learning-based software vulnerability detection requires high-quality datasets, which is essential for training effective models. To address challenges related to data label quality, diversity, and comprehensiveness, we constructed ICVul, a dataset emphasizing data quality and enriched with comprehensive metadata, including Vulnerability-Contributing Commits (VCCs). We began by filtering Common Vulnerabilities and Exposures from the NVD, retaining only those linked to GitHub fix commits. Then we extracted functions and files along with relevant metadata from these commits and used the SZZ algorithm to trace VCCs. To further enhance label reliability, we developed the ESC (Eliminate Suspicious Commit) technique, ensuring credible data labels. The dataset is stored in a relational-like database for improved usability and data integrity. Both ICVul and its construction framework are publicly accessible on GitHub, supporting research in related field.
\end{abstract}


\section{Introduction}
Detecting and mitigating software vulnerabilities remains one of the most critical challenges in ensuring the security and integrity of modern software systems. With the growing complexity of applications and their codebases, traditional methods of vulnerability detection, such as manual code inspection and static analysis~\cite{chakraborty2021deep}, struggle to keep up with the scale and intricacy of vulnerabilities present in large software projects. The need for automated, efficient, and accurate vulnerability detection has led to the rise of Machine Learning (ML) and Deep Learning (DL) techniques, which have shown promise in identifying potential security flaws by learning patterns from large volumes of code data~\cite{croft23}. However, the success of these models is heavily dependent on the quality, diversity, and comprehensiveness of the datasets used for training. Existing datasets often face challenges related to low label quality, small feature set, and poor balance. For instance, datasets such as BigVul~\cite{fan2020ac} and MegaVul~\cite{ni2024megavul} lack essential metadata about vulnerabilities, with vulnerable functions accounting for only 5\% of the total functions. Moreover, almost no datasets attempt to address noisy data in Fix Commits (FC) or incorporate Vulnerability-Contributing Commits (VCCs) as auxiliary information to improve vulnerability detection efforts.

To overcome these limitations, we introduce \textbf{ICVul}, an \textbf{I}ntegrated and \textbf{C}omprehensive \textbf{C}/C++ \textbf{Vul}nerability dataset. Unlike previous datasets that prioritize quantity, ICVul emphasizes data quality by offering a rich set of metadata, ensuring better balance, and excluding noisy data. Another key feature of ICVul is its inclusion of VCCs, which trace specific commits responsible for introducing vulnerabilities into the codebase using the state-of-the-art SZZ algorithm~\cite{iannone2022secret}. ICVul incorporates the ESC (Eliminate Suspicious Commit) technique, further enhancing label reliability by removing suspicious or ambiguous commits that could potentially mislead training processes. 

Our dataset not only supports traditional vulnerability detection tasks but also paves the way for more complex research in areas such as vulnerability prediction, risk assessment, and code quality analysis. The comprehensive metadata, along with careful data selection and labeling, ensures the dataset's reliability and suitability for training more accurate and robust models. Furthermore, ICVul enables the development of advanced models capable of handling multi-dimensional, multi-type data, and integrating with techniques like Just-in-Time (JIT) models~\cite{zhao2023systematic}, which are trained on VCCs data. Overall, the contributions of this paper are as follows:
\begin{itemize}
\item We present a well-labeled, well-balanced (class distribution), and integrated dataset for vulnerabilities in C/C++ projects. Our dataset goes beyond simply including vulnerable code in the FC. It also incorporates information from relevant VCCs, filters out noisy data (such as irrelevant file types), and employs the ESC technique to exclude unreliable commits. 
\item We introduce a vulnerability collection framework that can be re-run at any time to ensure the dataset remains up-to-date. The project, along with supporting scripts, is publicly available on GitHub\footnote{\url{https://github.com/Chaomeng-Lu/ICVul.git}}. 
\end{itemize}

\section{Dataset Construction}

\begin{figure*}[htbp]
    \centering
    \includegraphics[width=1\linewidth]{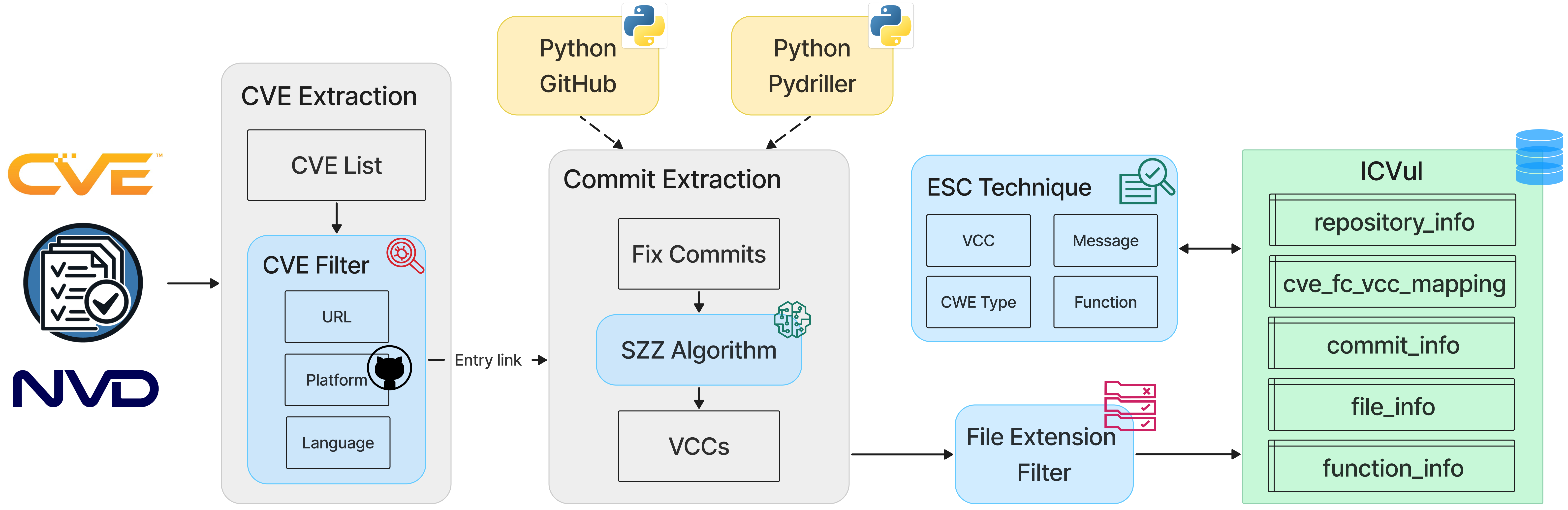}
    \caption{Overview of the ICVul construction framework.}
    \label{fig:Overview}
\end{figure*}

This section presents the key steps involved in constructing the dataset. It covers the extraction and filtering of CVE data, collection of VCC, extraction of commit, file, and function details, and the eliminate suspicious commit process to ensure label quality. Fig. \ref{fig:Overview} provides an overview of the ICVul construction framework.

\paragraph{CVE extraction and filtering}
The dataset construction process began with the collection of 269,509 CVEs, gathered on November 13, 2024. Initially, the analysis revealed a total of 924,834 reference URLs distributed across 11,370 different domains, with 66,059 URLs pointing to GitHub. The first step in data cleaning was to drop CVEs that lacked reference URLs or a Common Weakness Enumeration (CWE) ID, reducing the dataset to 248,505 entries. Next, CVEs without commit URLs (reference URLs contained keywords like ``/commit/") were removed, further narrowing the dataset to 13,733 records. Finally, the dataset was refined to include only CVEs associated with C/C++ repositories, reducing the number to 4,723, contains 5,366 potential FCs.

\paragraph{Vulnerability-Contributing Commits Collection}
After identifying the fixing commits, we extract the lines that were added or removed from the changed hunks (blocks of changes). These lines are then used to identify the blame commit, or VCC, which introduced the vulnerable code. For removed lines, we collect the blame commits where these lines were originally introduced. If no lines are removed in the fixing commit, we focus on the added lines and identify the blame for related contextual lines, such as those before and after the change. For example, large blocks of lines might represent an \textit{if} statement added as a safeguard to prevent a vulnerability. To trace the VCC, we employ the SZZ algorithm, as recommended by Iannone et al.~\cite{iannone2022secret}. The SZZ algorithm works by identifying the origin of the code protected by the newly added \textit{if} statement in the fix commit, thereby determining when the vulnerable code was first introduced or last modified.

\begin{figure*}[htbp]
    \centering
    \includegraphics[width=1\linewidth]{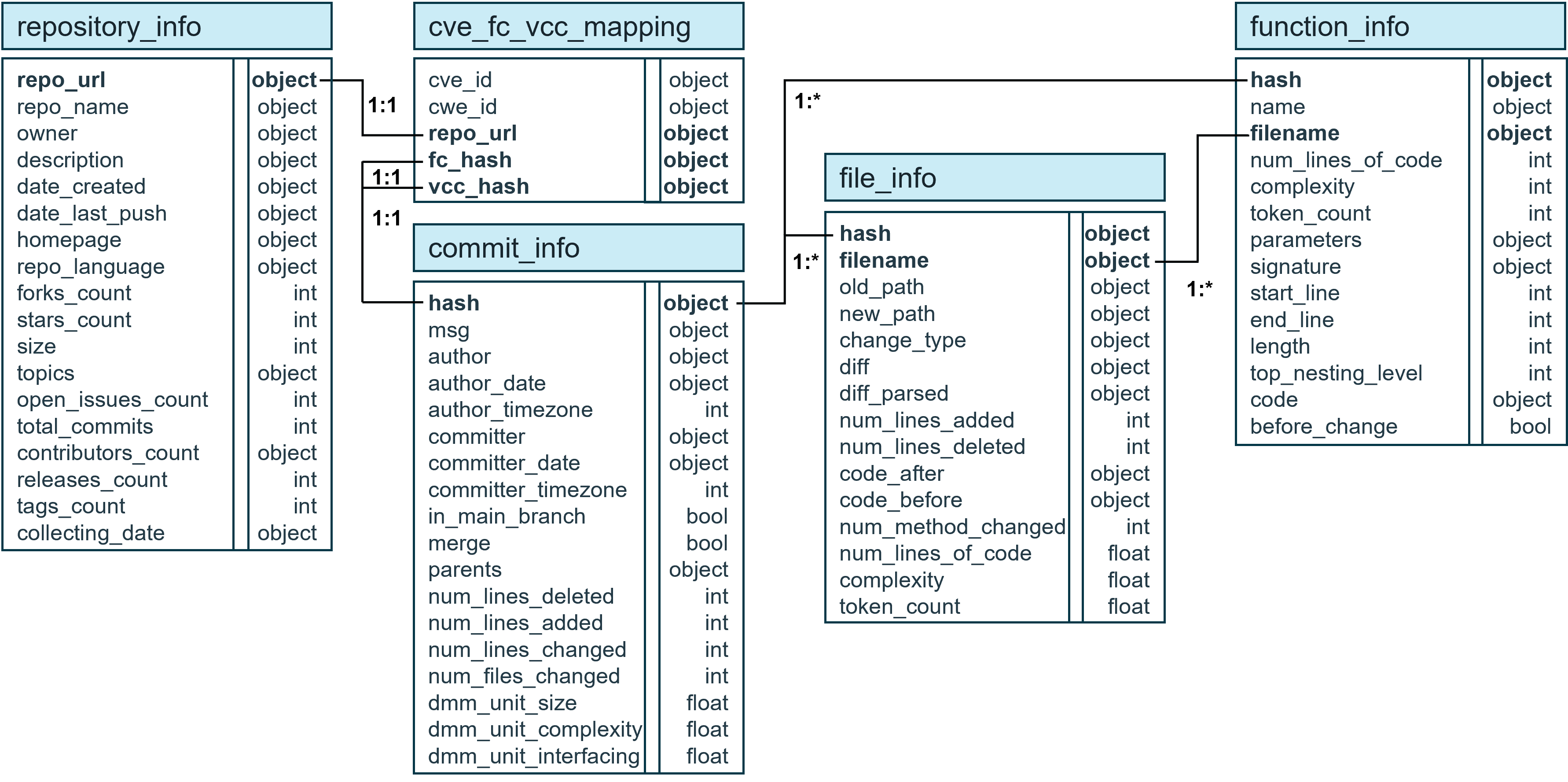}
    \caption{Structure of our ICVul Dataset.}
    \label{fig:DatasetS}
\end{figure*}

\paragraph{Commits/Files/Functions/Repositories Extraction}
At this stage, we first extract all the metadata related to repositories. Then, we begin by extracting all metadata associated with both the FCs and VCCs. For each FC, we proceed with file-level extraction, which includes capturing the metadata of the modified files as well as the file itself. This involves storing the versions of the files before and after the fix, labeled as code\_before and code\_after, respectively. Following the file-level extraction, we perform function-level extraction. This step captures metadata for all altered functions, including those added or deleted during the fix process. Each version of the function is stored separately and is appropriately labeled to indicate its state (before or after the fix). To minimize noise in the extraction process, we exclude any data, at both the file and function levels, that does not match the specified file extensions: \textit{.c}, \textit{.h}, \textit{.ec}, \textit{.ecp}, \textit{.pgc}, \textit{.cpp}, \textit{.cxx}, \textit{.cc}, \textit{.pcc}, \textit{.hpp}. This approach aligns with the methodology recommended by static analysis tool Flawfinder~\cite{Flawfinder}.

\paragraph{Eliminate Suspicious Commit}
After extracting the data, we address a key limitation in previous work that overlooks the noise present in FCs obtained from the CVE list, which are often reported to the NVD by different programmers across various projects. To address this, we developed the ESC technique for automatically screening noisy FCs. The ESC technique consists of four steps: 
\begin{enumerate}
    \item If the FC is blamed as a VCC of another FC, it is considered suspicious.
    \item If the FC belongs to more than one CWE type, it is considered suspicious.
    \item If the commit message of the FC is unclear (e.g. it contains keywords `merge'), it is considered suspicious.
    \item If the FC contains an unusually high number of vulnerable functions, it is considered suspicious. 
\end{enumerate}
In Step 1, we identified 283 suspicious FCs that may introduce a new vulnerability. In Step 2, we focused on ensuring accurate labeling at the CWE type level, identifying 98 suspicious FCs. In Step 3, we excluded merge commits and those unrelated to vulnerabilities, resulting in 81 suspicious FCs. In Step 4, we filtered out commits containing an unusually high number (set as 100, which is the top 0.02\%) of vulnerable functions by calculating the total number of vulnerable functions per commit, identifying 10 suspicious FCs. Each step independently addresses a different source of noise, effectively reducing false positives in the labeling process.

This approach helps eliminate 462 outliers in total (with some overlap between the four steps), accounting for 9.6\% of all reachable FCs. In terms of noise in functions, the ESC technique effectively filters out 23.8\% of the total functions (4,813 functions), including 24.2\% of the vulnerable functions (2,003 functions).

\section{Dataset Description}

ICVul  comprises five interconnected tables: repository\_info, cve\_fc\_vcc\_mapping, commit\_info, file\_info, and function\_info. As shown in Fig. \ref{fig:DatasetS}, inspired by CVEFixes~\cite{bhandari2021cvefixes}, the schema is designed to support a detailed analysis of software development and its associated vulnerabilities. Each table focuses on a distinct aspect of the data, providing detailed insights into repositories, commit histories, file-level modifications, and function-level characteristics. Specifically, the repository\_info table comprises 807 rows, providing detailed metadata about repositories. This includes information such as creation dates, primary programming languages, and popularity metrics like the number of stars and forks. The cve\_fc\_vcc\_mapping table maps CVE and CWE to repository and associated commit hashes, contains 4,712 records. The commit\_info table tracks commit-level details, including metadata such as authorship, lines of code added or deleted, and complexity metrics, contains 13,024 rows. The file\_info table captures file-level changes, while the function\_info table provides granular data on function complexity, token count, etc. The combination of these tables captures almost all information available in a software repository relevant to vulnerability detection in a data science-ready format.

\begin{figure}[htbp]
    \centering
    \includegraphics[width=1\linewidth]{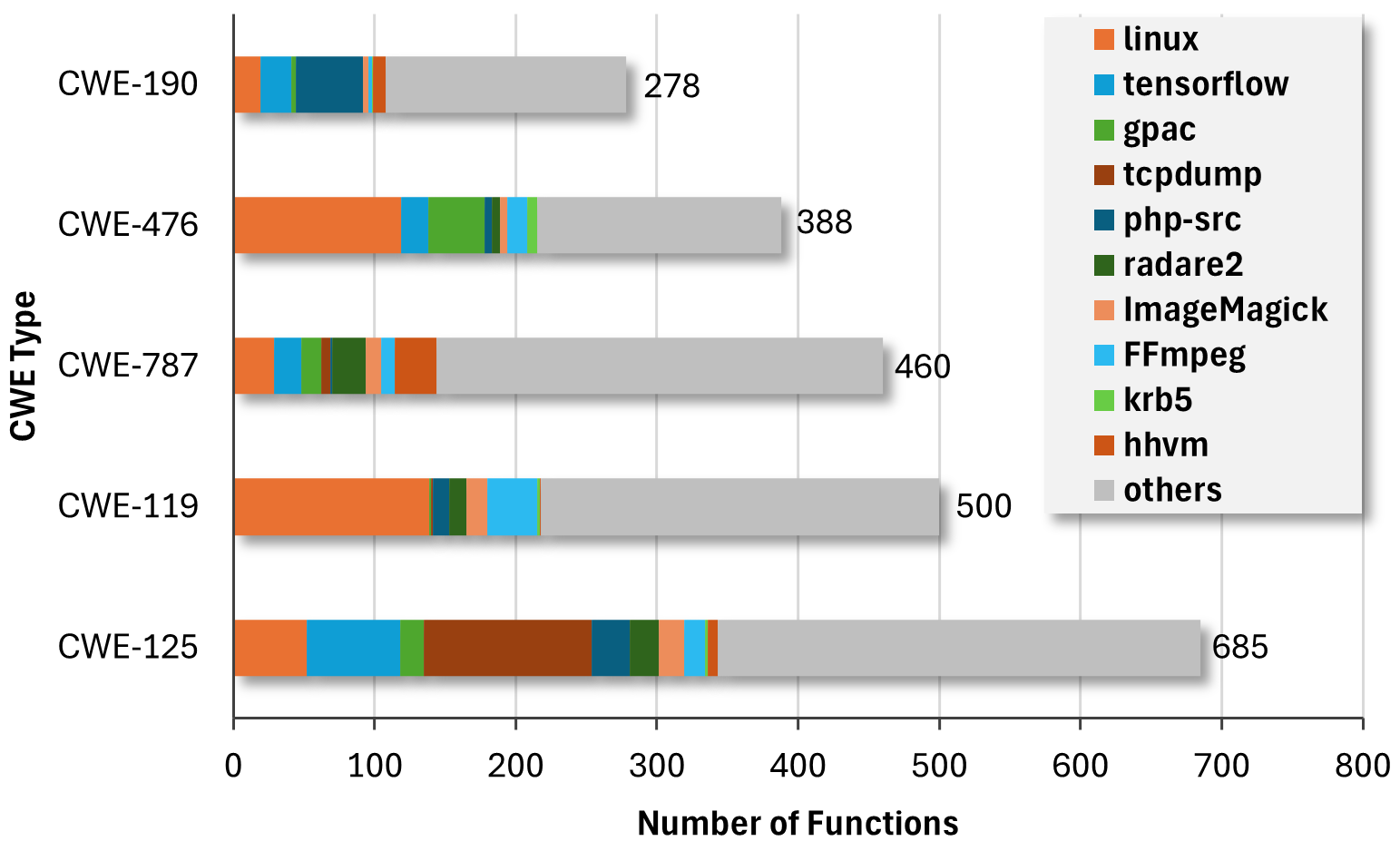}
    \caption{Distribution of top 10 repositories across the top 5 CWE types, sorted by vulnerable function count.}
    \label{fig:DistributionF}
\end{figure}

Fig. \ref{fig:DistributionF} shows the distribution of top 10 repositories, sorted by the number of vulnerable functions, across the top five CWE types based on function count. Table \ref{tab:comparisondataset} provides a comparative analysis of ICVul and other existing datasets used for studying vulnerabilities in software. It lists key attributes such as the number of repositories, CWEs, commits, files, functions, and vulnerable functions, along with the ratio of vulnerable functions to total functions. The table also highlights whether the datasets include VCCs, additional metadata, and whether they exclude noise. Since GitHub contains relatively comprehensive metadata and more standardized project management, we collect vulnerability information solely from this open-source platform. This results in our dataset having a smaller volume compared to other datasets. Nonetheless, the size of our dataset is still sufficient for training ML and DL models~\cite{steenhoek23}. Furthermore, by only collecting functions with changes in the commits to ensure label accuracy, the dataset achieves a much better balance ratio of 41\% at the function level. In addition, our dataset includes comprehensive metadata, VCC inclusion, clear CWE types, and noise exclusion, which contribute to higher label quality and more comprehensive data coverage.

\begin{table*}[htbp]
\centering
\caption{Comparative analysis of ICVul and existing datasets.}
\label{tab:comparisondataset}
\begin{tabular}{p{1.7cm}|r|r|r|r|r|r|p{1.2cm}|p{1.05cm}|p{1.1cm}|p{1cm}}
    \hline
    Dataset & Repositories & CWEs & Fix Commits & Files & Functions & Vul Funcs & Ratio of Vul Funcs & VCCs Inclusion & Metadata Inclusion & Noise Exclude\\
    \hline
    BigVul~\cite{fan2020ac} & 348 & 97 & 3,754 & 8,143 & 264,919 & 11,823 & 4\% & NO & NO & NO\\
    CrossVul~\cite{nikitopoulos2021crossvul} & 498 & 107 & 3,009 & N/A & 134,126 & 6,884 & 5\% & NO & NO & NO\\
    CVEFixes~\cite{bhandari2021cvefixes} & 564 & 127 & 3,641 & N/A & 168,089 & 8,932 & 5\% & NO & YES & NO\\
    DiverseVul~\cite{chen2023diversevul} & 797 & 150 & 7,514 & N/A & 330,492 & 18,945 & 6\% & NO & NO & NO\\
    MegaVul~\cite{ni2024megavul} & 992 & 169 & 9,019 & N/A & 322,168 & 17,380 & 5\% & NO & NO & YES\\
    \textbf{ICVul} & 807 & 146 & 4,327 & 6,862 & 15,396 & 6,276 & 41\% & YES & YES & YES\\
    \hline
    \multicolumn{11}{p{16.2cm}}{\small \textit{Note: CrossVul and CVEFixes are cross-language datasets, the statistics used for comparison in this table include only the data related to C/C++.}}
\end{tabular}
\end{table*}

\section{Data Applications}
ICVul serves as a robust resource for advancing research in software vulnerability detection (SVD) and related fields. 

\paragraph{Advancing Machine Learning for Vulnerability Detection}
ICVul provides a high-quality dataset tailored for training ML models to detect software vulnerabilities at the commit, function and file levels. With its extensive metadata, such as VCCs, code changes, and contextual details, researchers can develop complex models that accurately identify vulnerable patterns in code. The dataset’s focus on data quality, achieved through techniques like the ESC method, ensures that the models are trained on well-labeled data, improving their detection accuracy and reducing false positives. The dataset also enables the development of more complex models, such as those capable of learning from multi-dimensional, multi-type data or models integrated with JIT techniques. Additionally, the dataset provides clear vulnerability labels at the CWE type level, enabling the development and research of multi-class classification models.

\paragraph{Supporting Code Analysis and Security Research}
ICVul facilitates in-depth analysis of code behavior and its association with vulnerabilities. Researchers can leverage the dataset to study how specific code changes and patterns contribute to vulnerabilities, enabling a better understanding of software weaknesses. Additionally, the dataset’s relational-like structure promotes efficient exploration and manipulation, making it ideal for tasks like code quality assessment, secure development practice evaluation, and analyzing trends in vulnerability-prone code.

\section{Limitations}

ICVul, while emphasizing data quality, has certain limitations that should be acknowledged. Firstly, the dataset prioritizes quality over quantity, resulting in a smaller data volume compared to other datasets. Secondly, it exclusively focuses on open-source projects from GitHub, as these repositories provide relatively detailed information, which also limits the dataset's size and diversity. Thirdly, the SZZ algorithm used in this work may have certain limitations in terms of accuracy~\cite{herbold2022problems}. Lastly, ICVul does not include all potentially valuable metadata. For instance, at the function level, metadata such as Fan-in and Fan-out\footnote{Fan-in refers to the number of functions or modules that call a particular function, indicating its potential criticality or reusability. Fan-out measures the number of functions or modules that a particular function calls, reflecting its complexity or dependencies.} could provide additional insights, but these were excluded due to the immaturity of techniques required to reliably extract such information. These limitations highlight areas for potential future enhancements as methods and resources improve.

\section{Related work}

Early SVD model training often relied on synthetically generated datasets like SATE IV Juliet~\cite{okun2013report} and SARD~\cite{sard2024}, which are limited by predefined vulnerability patterns and do not fully reflect real-world complexity. Then, Zheng et al.~\cite{zheng2021d2a} introduced the D2A Dataset, using differential analysis to label vulnerabilities identified by static analysis tools, though its label quality remains uncertain due to the low accuracy of static analyzers. Manually labeled datasets based on real repositories like Devign~\cite{zhou-2019} and Reveal~\cite{chakraborty2021deep} face challenges such as high labor costs, limited data, and lack of project diversity~\cite{croft23}. To address these, automated datasets like BigVul~\cite{fan2020ac} were developed. Similarly, CrossVul~\cite{nikitopoulos2021crossvul} and CVEfixes~\cite{bhandari2021cvefixes} also collect vulnerable code and patches from open-source repositories. CrossVul gathers data from Git commits, while CVEfixes pulls CVE records from the NVD. DiverseVul~\cite{chen2023diversevul} offers the largest C/C++ dataset by extending beyond NVD records. MegaVul~\cite{ni2024megavul} implements certain measures to exclude noisy data. However, their measures do not extend to the commit level. Consequently, this oversight may result in the inclusion of merge commits or commits with unclear intent.

\section{Conclusion}
This paper introduces ICVul, a high-quality dataset designed to advance ML-based SVD by addressing challenges related to label quality, diversity, and comprehensiveness. Through a meticulous construction process, including the use of the SZZ algorithm to trace VCCs and the ESC technique for reliable labeling, ICVul provides clean and credible data. The dataset, stored in a relational-like structure, enables efficient analysis and supports various applications, including vulnerability detection, secure software development. As future work, we will focus on expanding the dataset by incorporating a broader range of programming languages and platforms, enhancing its diversity.

\clearpage
\balance
\bibliographystyle{IEEEtran}
\bibliography{reference}

\end{document}